\documentclass{llncs}
\usepackage[dvips]{graphics}
\usepackage{graphicx}
\usepackage{amsmath}
\usepackage{amssymb}
\usepackage{amsfonts}
\usepackage{txfonts}
\usepackage{pxfonts}
\usepackage{xcolor}
\usepackage{multirow}
\usepackage{array}
\usepackage{float}
\usepackage{colortbl}
\usepackage{arydshln}
\usepackage{slashbox}
\usepackage{hyperref}

\newcommand{\getsr}{\stackrel{{\scriptscriptstyle\$}}{\gets}}

\begin{document}

\title{Towards Blockchain-enabled Searchable Encryption}
\author{Qiang Tang}

\institute{Luxembourg Institute of Science and Technology\\
5 Avenue des Hauts-Fourneaux\\
4362 Esch-sur-Alzette, Luxembourg\\
\email{qiang.tang@list.lu}}

\maketitle

\centerline{\today}

\begin{abstract}
Distributed Leger Technologies (DLTs), most notably Blockchain technologies, bring decentralised platforms which eliminate a single trusted third party and avoid the notorious single point of failure vulnerability. Since Nakamoto's Bitcoin cryptocurrency system, an enormous number of decentralised applications have been proposed on top of these technologies, aiming at more transparency and trustworthiness than their traditional counterparts. These applications spread over a lot of areas, e.g. financial services, healthcare, transportation, supply chain management, and cloud computing. While Blockchain brings transparency and decentralised trust intuitively due to the consensus of a (very large) group of nodes (or, miners), it introduces very subtle implications for other desirable properties such as privacy. In this work, we demonstrate these subtle implications for Blockchain-based searchable encryption solutions, which are one specific use case of cloud computing services. These solutions rely on Blockchain to achieve both the standard privacy property and the new fairness property, which requires that search operations are carried out faithfully and are rewarded accordingly. We show that directly replacing the server in an existing searchable encryption solution with a Blockchain will cause undesirable operational cost, privacy loss, and security vulnerabilities. The analysis results indicate that a dedicated server is still needed to achieve the desired privacy guarantee. To this end, we propose two frameworks which can be instantiated based on most existing searchable encryption schemes. Through analysing these two frameworks, we affirmatively show that a carefully engineered Blockchain-based solution can achieve the desired fairness property while preserving the privacy guarantee of the original searchable encryption scheme simultaneously.
\end{abstract}

\section{Introduction}

With the prevalence of cloud computing, many organizations are outsourcing their data and services to the cloud. By doing so, an organization or individual can enjoy a wide spectrum of benefits such as agileness and cost-saving. Moreover, the cloud service provider can deploy sophisticated cybersecurity solutions to meet the requirements from the relevant security regulations. It is widely perceived that the big cloud service providers, such as Amazon and Microsoft, provide better protection in practice than most organizations if they do it by themselves. However, there are indeed drawbacks for the outsourced data and corresponding operations, among which \emph{loss of privacy} is the most significant one. As the cloud service provider can observe the data usage patterns and potentially have access to the plain data, it becomes a concern when the data contains sensitive information. The issue becomes more complex when the services are cross-border and need to comply with privacy regulations from different regimes. Besides privacy, the verifiability of outsourced computing tasks might also be a serious concern. In order to save cost, the cloud service provider might not carry out the promised tasks faithfully. At the end, the incomplete or even flawed computing results might damage the client's business severely. Therefore, how to guarantee privacy and verifiability in outsourcing has been an active research area for many years.

Regarding the potential computing tasks on outsourced data, search is the most fundamental one. To cater to the privacy needs, searchable encryption is a category of cryptographic primitives that allows data to be outsourced in an encrypted form while still being able to be searched over. Searchable encryption typically assumes a standard client-server setting, where a client outsources its encrypted data to a cloud server, which can then search on the client's behalf without decrypting the data. Existing searchable encryption schemes can be broadly classified into two settings. In the asymmetric setting \cite{boneh:1}, the client can publish a public key, by which anybody can generate searchable encrypted data and store it on the server. Later, the client, who has access to the private key, can let the server search on its behalf by issuing a trapdoor. In the symmetric setting \cite{song:1}, a client uses symmetric keys to encrypt its own data and stores the ciphertexts on the server. Later on, as in the asymmetric setting, the client can let the server search on its behalf by issuing a trapdoor. Compared to the symmetric setting, the asymmetric setting poses higher challenges to data privacy, as shown in \cite{tang:65}. This implies that asymmetric searchable encryption schemes leak much more information in reality, and potentially make such schemes very undesirable facing strong attackers.

In this paper, we focus on symmetric searchable encryption schemes and show how to properly leverage Blockchain to achieve verifiability and more properties without sacrificing privacy. Nevertheless, our discussions and approaches could also be applied to other tasks.

\subsection{Preliminary on Symmetric Searchable Encryption}
\label{pre}

We assume the client has a database $\mathcal{DB}$, which contains the files which will be searched based on an inverted inverted index. We assume a basic version of symmetric searchable encryption scheme which only consists of two stages $\mathsf{Setup}$ and $\mathsf{Search}$, with an example shown in Section \ref{sec:huscheme}. In the $\mathsf{Setup}$ stage, the client extracts a keyword set $\mathcal{W}$ from the files in the $\mathcal{DB}$ and builds an encrypted inverted index, which is then stored on the server. In the $\mathsf{Search}$ stage, the client interacts with the server to search for the files which contain any keyword $w \in \mathcal{W}$.

\begin{remark}
Since search is done over an inverted index, the actual files are not relevant anymore once the inverted index has been constructed. Therefore, the protection of the real files from $\mathcal{DB}$ can be done separately (e.g. they can be encrypted via standard symmetric encryption schemes and stored on the same or a different server). As a result, the search will result in file identifiers instead of real files. Some schemes support additional operations, such as add, delete or update the encrypted index, while others enable more complex search queries such as conjuncted keywords. We leave them out for simplicity reasons. Nevertheless, extending our discussions to these more general schemes will be an interesting and separate line of research work.
\end{remark}

To facilitate our discussions, we provide a high-level workflow of both stages. Existing schemes might optimize their performances or security with very specific tricks, e.g. index data structure. Nevertheless, most of them follow the workflow. In the $\mathsf{Setup}$ stage, the following operations will occur.

\begin{enumerate}
    \item Run by the client, it first generates the key materials, namely private key(s).
    \item Given the database $\mathcal{DB}$, a keyword set $\mathcal{W}$ is extracted and a plaintext inverted index is built. The index is informally a table, shown in Figure \ref{fig:inverted} for example, where each row contains the file identifiers associated with a specific keyword.  Note that the client might choose to pad the rows so that they contain certain number of file identifiers, e.g. all rows can be padded to contain the same number of file identifiers.
        \begin{figure}[h]
        \centering
        \includegraphics[scale=0.68]{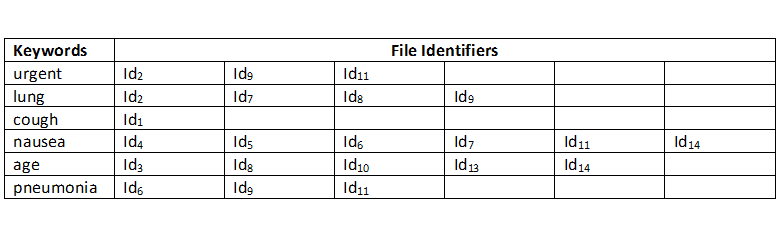}
        \caption{Inverted Index Example}
        \label{fig:inverted}
        \end{figure}
    \item Using the private key(s), the client \emph{encrypts} the inverted index and obtains an encrypted form of it. Note that the \emph{encryption} here means not only the hiding of keyword and identifier information but also can be the hiding of other pattern information such as the ordering of the encrypted keywords and identifiers.
        Finally, the encrypted index is stored on the server.
\end{enumerate}

In the $\mathsf{Search}$ stage, if the client wants to find all the file identifiers associated with a keyword $w$, the following operations will occur.

\begin{enumerate}
    \item Using the private key(s), the client generates a trapdoor $T_w$ based on its private keys(s) and $w$, and sends it to the server.
    \item With the trapdoor $T_w$, the server can go through the encrypted index and match those elements which contain the same keyword (i.e. $w$) as that embedded in the trapdoor.
    \item For the matched elements, the server recovers the associated file identifiers, denoted as a set $\mathcal{ID}_w$ and return them to the client.
\end{enumerate}


\subsection{Privacy and Fairness Challenges}
\label{properties}

Searchable encryption can be seen as a derivative of standard encryption primitives, but it is more complex due to the fact that, concerning privacy, we need to consider not only the encrypted index but also the trapdoors. If there is a secure channel between the client and the server, then the server is the main privacy attacker. Intuitively, we will expect at least the encrypted index or trapdoors alone do not leak any information about the embedded keywords. This can be formulated in a similar way to the semantic security property of encryption schemes \cite{goldwasser84}, and easy to achieve. However, the situation is more complex for searchable encryption, due to the fact that search operations link the encrypted index and trapdoors so that more information will be leaked. In more detail, there are concerns of access pattern leakage and search pattern leakage.
\begin{itemize}
    \item Informally, access pattern is the file identifier information resulted from the client's search queries.
    \item While, search pattern is about whether two trapdoors contain the same keywords or not.
\end{itemize}
These two types of leakages are clearly closely related. After receiving several trapdoors, even if the server might not learn the keywords, it can derive statistical information about the keywords based on access pattern. In practice, the statistical information can disclose the search pattern and even lead to full recovery of the keywords. Besides privacy, the other practical concern related to searchable encryption is the verifiability of search results. As we have mentioned before, it is desirable for the server to assure the client that the search results have come from a faithful execution of the protocol. On the other side, it is also desirable that the server is rewarded properly for the faithful execution of the search protocol. Follow the literature work, if a searchable encryption solution satisfies both requirements simultaneously, we say it is \emph{fair}.

So far, very little has been done to design privacy-preserving and fair searchable encryption solutions, except for some recent solutions that leverage on Blockchain to achieve fairness \cite{chen19,Hu}. We note that there are verifiable symmetric searchable encryption schemes, e.g. \cite{ChaiG12}, which however only guarantee that a semi-honest server will follow the protocol. Being a technology that brings \emph{trust} as many believe, Blockchain infact results in very subtle tradeoffs among the desirable properties, e.g. privacy and verifiability. Unfortunately, these tradeoffs have been ignored by many researchers. As such, it remains as an open question how well these recent searchable encryption solutions have addressed the privacy and fairness requirements.

\subsection{Contribution and Organisation}

Our contribution in this paper is two-fold. We start by examining some recent Blockchain-based searchable encryption solutions, i.e. \cite{chen19,Hu}. We show that directly replacing the server of a searchable encryption scheme with a Blockchain is a very undesirable solution. First of all, it introduces considerable cost with respect to storing the encrypted index and executing the smart contract which implements the search operation. Secondly, these solutions suffer from the inherent issues of Blockchain, e.g. the forking problem\footnote{\url{https://en.wikipedia.org/wiki/Fork_(blockchain)}}. This might cause serious usability issues for these solutions. Thirdly, the privacy concerns of the underlying scheme are amplified by the Blockchain platform. The access pattern and search pattern leakages are exposed to all entities who can access the Blockchain. To mitigate the identified issues in our analysis, we then propose two frameworks that can be instantiated based on most existing searchable encryption schemes. In both frameworks, search operations are carried out by the server(s) as in the traditional schemes, while Blockchain is leveraged to achieve the fairness property only.

The rest of this paper is organised as follows. In Section \ref{blockchain}, we give a brief summary to Blockchain technologies. In Section \ref{sec:dltsol}, we present and analyse the existing Blockchain-based solutions. In Section \ref{newdesign}, we present our new frameworks and provide corresponding analysis. In Section \ref{con}, we conclude the paper.

\section{Blockchain in a Nutshell}
\label{blockchain}

Since the seminal report from Nakamoto \cite{bitcoin}, the concept of Blockchain has become very popular not only in the research community but also in the society at large. Its popularity largely comes from the fact that it is the key enabling technology for the variety of cryptocurrency systems, including Bitcoin \footnote{\url{https://bitcoin.org/en/}} and the altcoins, even though the history of both the idea of cryptocurrency and the techniques in Blockchain can be traced back to much earlier era \cite{Narayanan}. As a matter of fact, today there are over 1600 such systems according to Wikipedia \footnote{\url{https://en.wikipedia.org/wiki/List_of_cryptocurrencies}}.

\subsection{Blockchain Overview}

\begin{figure}[h]
\centering
\includegraphics[scale=0.48]{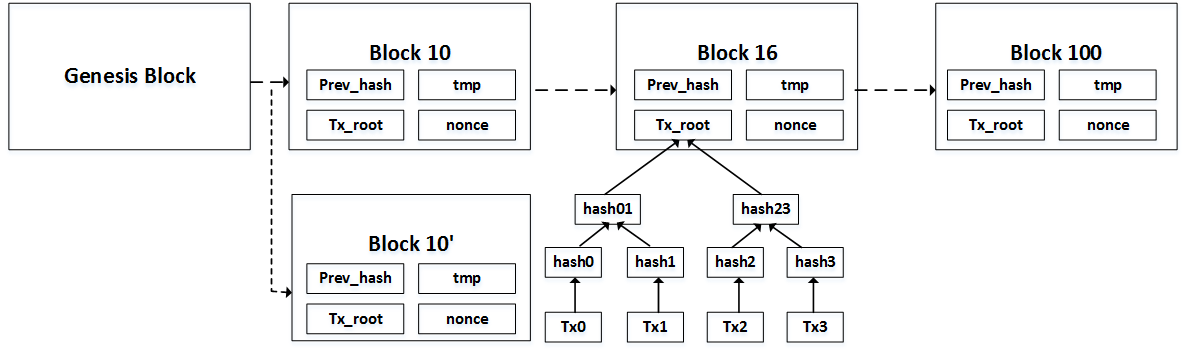}
\caption{Blockchain Structure}
\label{fig:blockchain}
\end{figure}

Informally, the data on a Blockchain is organized in the form shown in Figure \ref{fig:blockchain}.
Depending on who maintains the chain (i.e. generate and approve new blocks), Blockchain systems can be roughly divided into two categories. If anyone can publish and approve a new block, it is permissionless. Otherwise, if only particular nodes are allowed to do it, it is permissioned. More details about the categorisation can be found in the NIST report \cite{nist}. From now on, we refer to these particular privileged nodes as \emph{miners} in the paper.

As the core characteristic of Blockchain, repeatedly, a certain number of new data entries (e.g. transactions) will be packed into a new block and appended to the existing (longest) chain. In the case of Bitcoin Blockchain, a new block also includes the hash value of the last block of the current chain. The block is formed with some specific features, e.g. a proof of work needs to be carried out so that the hash value of the new block contains some number of consecutive zeros. The new block will be broadcast to the whole network, and it will be accepted in the network after everything being validated. Depending on the variants and implementations, there are many subtle details on how a block is formed and accepted to the chain, we refer the readers to the corresponding technical specifications for the precise information.

Besides cryptocurrencies, Blockchain systems act as the key foundation platform for \emph{smart contracts}, which facilitate automated execution of software programs in a verifiable manner. One of the notable examples is Ethereum \footnote{\url{https://www.ethereum.org/}}, which is the second largest cryptocurrency system after Bitcoin and gains the popularity because of its powerful smart contracts functionality.  In practice, smart contracts can enable a variety of trustworthy distributed applications, e.g. building digital Decentralized Autonomous Organizations (DAOs).

It is worth noting that Blockchain represents one special case of the broader distributed ledger technologies (DLTs), which are decentralised databases that rely on independent computers to record, share and synchronize digital transactions. In many cases, say IOTA \footnote{\url{https://www.iota.org/}} and Hyperledger \footnote{\url{https://www.hyperledger.org/}}, the decentralised database is not organised in a single chain of blocks. Instead, it can be a graph (e.g. IOTA) or multiple chains of blocks (e.g. Hyperledger). Despite the different forms, a DLT can possess similar properties to those from a Blockchain. For a more comprehensive review of cryptocurrencies, Blockchain and DLT technologies, we refer the readers to the comprehensive books such as \cite{diedrich,Narayanan1,swan}.

\subsection{Properties of Blockchain}
\label{subsec: properties}

Regardless of the forms of a Blockchain or DLT in general, the following useful properties can be expected.
\begin{itemize}
\item \emph{Democracy and Decentralised Control.} Everyone can potentially act as a miner and has the same privilege to generate blocks and approve blocks to the Blockchain. This is generally true for systems employing the proof of work (PoW) as the consensus mechanism in the permissionless scenario, while it can be different in other cases. Regardless, Blockchain eliminates a single fully trusted party and avoids the single point of failure vulnerability correspondingly.
\item \emph{Integrity and Immutability.} If an attacker or a group of colluded attackers does not dominate the consensus process, e.g. in the case of Bitcoin Blockchain more than 51\% of the computing power is at the hands of semi-honest miners (see the explanation below for the semi-honest assumption), then it will not be able to modify the existing blocks that have been agreed on by the consensus.
\item \emph{Consistency.} There is a single consistent view of the chain even facing strong attackers, based on assumptions mentioned above. However, note that when nodes deviate from the predefined rules, forks could be generated and there will be different views from different players, e.g. in the case of Ethereum\footnote{\url{https://en.wikipedia.org/wiki/Ethereum}}.
\end{itemize}

These properties further provide certain levels of auditability and transparency, and generally increase the trustworthiness of the system. These aforementioned properties or even a subset of them can be very desirable for many applications from different sectors. Some people have considered  Blockchain as a trust machine for the society \footnote{\url{https://www.economist.com/leaders/2015/10/31/the-trust-machine}}. The trust that users have towards Blockchain systems is mainly from the fact that the majority of miners will be semi-honest from the cryptographic perspective. The semi-honest assumption basically says that these miners will follow the predefined protocols to perform what has been specified and programmed in the Blockchain software, and particularly this excludes the possibility that they will collude to interfere with the normal Blockchain operations. For PoW-based Blockchain, the trust depends on the common assumption that 51\% of the computing power lies at the hands of semi-honest miners. While for other types of Blockchain, corresponding assumptions need to be made. For example, for Proof of Stake (PoS)-based DLTs, we need to assume that the parties that possess the majority of stakes will behave honestly.

Besides cryptocurrencies, Blockchain has been widely promoted in designing decentralised protocols, e.g. fair secure multiparty computation protocols \cite{Choudhuri17},  confidentiality-preserving smart contracts \cite{Cheng19}, double auction \cite{Zavodovski19}, and the Blockchain-based searchable encryption schemes \cite{chen19,Hu}. In most of these works, Blockchain is treated as a trusted platform that achieves some of the aforementioned properties persistently. However, we observe a dilemma with this trust assumption and raise concerns about the feasibility of (some) existing solutions. Let's suppose a client originally deploys a service at a dedicated cloud server, and now it wants to leverage Blockchain to improve the security.

\begin{itemize}
\item On one hand, all the promises of a Blockchain come from the holy assumption that no single entity can significantly influence the operations of the system and everything should be based on a consensus \footnote{In fact, it is more complex when it comes to questions such as how the evolution of a Blockchain platform should proceed, see the case of Bitcoin Blockchain}. This means that a normal user like the client, will not play any significant role, particularly the client may not be able to determine the miners or even know them.

\item On the other hand, from the perspective of the client, it may desire absolute certainty regarding the status of the Blockchain, the promised properties, and other aspects such as efficiency and cost. Unfortunately, the satisfaction of these requirements will depend on the consensus of some entities, which are not supposed to be influenced by the client.

\end{itemize}

Clearly, there is a governance dilemma facing the client when it wants to deploy its service on Blockchain. More consequences of this dilemma can be found in Section \ref{subsec:anablockchain} and \ref{ana:privacy}. Nowadays, this dilemma is hindering the deployment of Blockchain-based services , e.g. see the IBM-Maersk case \footnote{\url{https://www.coindesk.com/ibm-blockchain-maersk-shipping-struggling}}.

\section{Blockchain-enabled Searchable Encryption}
\label{sec:dltsol}

In this section, we first briefly recap the Blockchain-based searchable solutions from \cite{chen19,Hu}, and then present our analysis results from the economic, security and privacy aspects.

\subsection{Description of the Existing Solutions}
\label{sec:huscheme}

The central idea of solutions from \cite{chen19,Hu} is to treat Blockchain (that supports smart contracts) as a transparent and neutral platform. Intuitively, these solutions just replace the server in traditional scenarios with a Blockchain, which interacts with the client via a smart contract. All search and fairness-related logics are programmed into the smart contract. Based on the transparency and neutrality assumptions, the following notion of "fairness" can be achieved: (1) search operations will be performed in the pre-defined manner if we assume that a majority of the miners will not collude with each other; (2) the miner(s) will be rewarded for their search operations due to the fact that deposits are required before any search operation is carried out.

Let the client's database be denoted as $\mathcal{DB}$. Next we review the solution from \cite{Hu}. For simplicity, we only review the \emph{Setup} and \emph{Search} stages, while skipping the \emph{add} and \emph{delete} stages as they do not affect our analysis.

\begin{itemize}

\item $\mathsf{Setup}(\mathcal{DB}, \lambda)$: run by the client, the following operations are performed.

\begin{enumerate}
\item Initialize an empty list $L$, an empty dictionary $\sigma$, a counter $c$, and a block size $p$.
\item Extract a keyword set $\mathcal{W}$ from $\mathcal{DB}$.
\item Select two pseudorandom functions $\mathsf{F}$ and $\mathsf{G}$; Generate a secret key $K \getsr \{0,1\}^{\lambda}$.
\item For every keyword $w \in \mathcal{W}$, do the following
    \begin{enumerate}
    \item Compute $K_1 =\mathsf{F}(K, 1||w)$ and $K_2 = \mathsf{F}(K, 2||w)$, where $||$ is a concatenation operator.
    \item Set $\alpha = \lfloor \frac{|\mathsf{DB}(w)|}{p}\rfloor$ and $c=0$, where $\mathsf{DB}(w)$ is the file identifier set associated with $w$ and $|\mathsf{DB}(w)|$ indicates the number of identifiers in the set.
    \item Divide $\mathsf{DB}(w)$ into $\alpha+1$ blocks, and pad the last block into $p$ entries if necessary.
    \item For each block in $\mathsf{DB}(w)$, do the following
        \begin{enumerate}
        \item Set $\tilde{id}=id_1||\cdots||id_p$, $r \getsr \{0,1\}^{\lambda}$, $d= \tilde{id} \oplus \mathsf{G}(K_2, r)$, $l=\mathsf{F}(K_1, c)$.
        \item Add $(l, d, r)$ to the list $L$ in lex order.
        \item Set $c=c+1$.
        \end{enumerate}
    \end{enumerate}

\item Set $EDB=L$, partition $EDB$ into $n$ blocks $EDB_i$ $(1 \leq i \leq n)$ and send them to the smart contract.
\item For each received $EDB_i$, the smart contract parses each entry in $EDB_i$ into $(l, d, r)$ and add it to the Blockchain.
\end{enumerate}

\vspace{0.2cm}

\item $\mathsf{Search}(K, w; *)$: run between the client and the Blockchain (via the smart contract), the following steps are followed.

    \begin{enumerate}
    \item The client computes $K_1 =\mathsf{F}(K, 1||w)$, $K_2 = \mathsf{F}(K, 2||w)$.
    \item The client sets $c=0$, and sets an iteration number $R$ and step size $step$.
    \item For $0 \leq i \leq R$, do the following
    \begin{enumerate}
    \item The client sets $ST_i=(K_1, K_2, c)$ and sends it to the smart contract.
    \item The smart contract asserts the gas cost is lower than the balance, and then performs the following steps for $i=0$ until $i \geq step$. Note that the solution assumes an Ethereum platform.
        \begin{enumerate}
            \item Set $\ell = \mathsf{F}(K_1, c)$.
            \item If $\mathsf{Get}(\ell)=\perp$ stop; otherwise, set the result to be $(d, r)$. The $\mathsf{Get}$ function simply retrieves the tuple with the same $\ell$.
            \item Compute $\tilde{id} = d \oplus \mathsf{G}(K_2, r)$;
            \item Parse and save $\tilde{id}$.
            \item Set $c=c+1$.
            \item Set $i=i+1$.
        \end{enumerate}
    \end{enumerate}
    \end{enumerate}

\end{itemize}

The solution from \cite{chen19} is pretty the same as the above solution. The main difference is that it assumes a specific electronic health record (EHR) application scenario and the keyword is in the form of an expression like "(disease = `disease name`) AND (num1 $\leq$ age $\leq$ num2)".

\subsection{General Analysis w.r.t. Blockchain Usage}
\label{subsec:anablockchain}

From an economic perspective, in comparison to a dedicated server based solution, it is clear that a Blockchain-based solution will incur more costs regarding storage and computations, because several miners will need to perform the same tasks in parallel. The computational cost might become a more significant concern if a PoW-based consensus is employed by the underlying Blockchain platform. In connection to \emph{fairness}, one hidden concern is about the cost model for the miners of the Blockchain. By default, it is common to estimate the cost of operations based on the computations incurred by the smart contract executions. However, the real cost for the miners goes beyond that. For example, there is also cost for the communication and storage. In addition, the miners need to guarantee their availability for the searchable encryption services, which means investment in security and diaster recovery countermeasures. In the proposed solutions \cite{chen19,Hu}, it remains as a question how the client should estimate these costs and include them in the offer. In connection to \emph{search complexity}, a potential concern is the storage of the encrypted index when addition and deletion are enabled\footnote{Note that such features are included in the solution by Hu et al. \cite{Hu}.}. Suppose the Blockchain serves for searchable encryption services for many clients, it will be the case that the newly added encrypted index for a specific client will be stored in block(s) which are far away from those storing the previous encrypted index. This means the search operation may need to traverse through the whole chain to cover all relevant indexes for the specific client. The situation gets much worse if a permissionless Blockchain is employed because other applications will add a tremendous number of blocks over the time.



Contrary to the common belief that Blockchain could act as a "trust" machine to build secure applications, it actually brings its inherent security risks that can be fatal to the applications on top. One prominent security issue is around smart contracts, where one well-known example is the  decentralized autonomous organization (DAO) attack in 2016\footnote{\url{https://en.wikipedia.org/wiki/The_DAO_(organization)}}, which has exploited some software bugs in the underlying Ethereum smart contracts, that leads to the transfer of 3.6 million Ether to the attacker's account. As a result of the attack, the Ethereum Blockchain had to make a hard fork due to the lack of a unanimous consensus on the solution. Besides smart contracts, Blockchain systems in general are also subject to other attacks, e.g. those against consensus mechanisms and distributed denial-of-service (DDoS) attacks \cite{blockchainattacks}. In the traditional setting, these issues might be easier to avoid or solve, or at least they can be solved much more quickly.

\subsection{Specific Analysis w.r.t. Privacy Guarantee}
\label{ana:privacy}

As noted in Section \ref{properties}, almost all searchable encryption schemes leak certain information to the server. Although a symmetric searchable encryption scheme leaks less than its asymmetric counterpart, the leakage might still be considered to be non-negligible. Take the scheme from Section \ref{sec:huscheme} as an example, there are (at least) two kinds of leakages, which are commonly shared by other similar schemes.

\begin{itemize}
\item \emph{search pattern leakage}. The $\mathsf{Search}$ algorithm is a deterministic function. This means that if the client searches the same keyword more than once, then the Blockchain miners will notice it. Based on such information, statistics such as frequency of searched keywords can be established. In turn, such statistics may allow the miners to recover the underlying keywords.

\item \emph{access pattern leakage}. In the $\mathsf{Setup}(\mathcal{DB})$ algorithm, $\mathsf{DB}(w)$ will be padded to guarantee that every block has exactly $p$ entries. However, this padding operation does not anonymize the index length very well. Let's assume $\mathsf{DB}(w_1)$ has $p+1$ entries and $\mathsf{DB}(w_2)$ has $2p+1$ entries. In this case, even after the padding, the keyword $w_2$ will result in $p$ more entries on the Blockchain than the keyword $w_1$. As a result, the $\mathsf{Search}$ operation may reveal the size relationship of the file identifier sets associated with the searched keywords.
\end{itemize}

Specific to this scheme, it is undesirable to reveal $\tilde{id}$ to the smart contract. Nevertheless, this can be easily resolved by not sending $K_2$ to the smart contract and instead the client decrypts $d$ by itself to recover $\tilde{id}$ at the end of the search operation.

In comparison to the traditional scenario without using a Blockchain, where the privacy information leakage is only limited to a single server, the leakage is amplified in Blockchain-based solutions. The degree of the amplification effect depends on which type of Blockchain is chosen by the solution.
\begin{itemize}
\item If a permissionless Blockchain is used, then the information leakage is available to everybody on the Internet. This is very likely not acceptable.

\item If a permissioned Blockchain is used, then the information leakage is provisionally available to those entities that can access the Blockchain. At the first glance, the privacy issue is less serious than the other case. However, the situation is rather complex, unless the searchable encryption scheme is deployed on a dedicated Blockchain where the right to read the Blockchain data is controlled by the client. Otherwise, the client will have no control over who will see its encrypted index and search history (e.g. no control over who will be the miners), which means it suffer from unexpected privacy leakage. This exemplifies the dilemma stated in Section \ref{subsec: properties}.
\end{itemize}

For searchable encryption applications, the immutability property of Blockchain might not be really necessary. In the contrary, this property might be undesirable concerning privacy protection. If the encrypted index and search histories live forever on a Blockchain, then it will stay as a persistent attack surface for any (emerging) attackers.

\subsection{Summary and Roadmap}

So far, we have analysed the advantages as well as disadvantages of Blockchain-based solutions. Our analysis indicates that a solution built directly based on a Blockchain, either permissioned or permissionless, causes issues from different aspects, including cost, security and privacy. For the studied solutions, it seems that the disadvantages will outweigh the advantages in practice. Nevertheless, this does not imply that Blockchain is useless for these applications, rather we believe that Blockchain will be a very useful tool to guarantee the fairness property. Without it, it will be a very sophisticated task to design fair and privacy-preserving searchable encryption solutions, and it may need to make significant changes to existing privacy-preserving searchable encryption schemes.

Based on our analysis, towards designing privacy-preserving and fair searchable encryption solutions, a modular approach seems more appropriate: exploiting Blockchain for the fairness guarantee and relying on dedicated server(s) for the actual search operations on the basis of an existing searchable encryption scheme. The key  challenge is to guarantee that the involvement of Blockchain does not affect the privacy guarantees of the underlying searchable encryption scheme. This leads to two new frameworks in the next section.

\section{New Generic Blockchain-based Frameworks}
\label{newdesign}

In this section, we propose two generic Blockchain-based frameworks, that can be instantiated based on most symmetric searchable encryption schemes. In both frameworks, there are three types of entities involved.

\begin{itemize}
\item \emph{Client}: The client is the party that wants to outsource its encrypted index. 

\item \emph{Server(s)}: As in the traditional setting, the server(s) store the encrypted index and carry out the search operations.

\item \emph{Blockchain}: The Blockchain acts as a semi-trusted platform to ensure fairness.


\end{itemize}

We assume there is a secure communication channel ( for confidentiality and integrity) between the client and all server(s), while there is no such a link between any entity and the Blockchain but we do assume that only the legitimate entity can communicate with the smart contract (which means a channel with integrity protection only).

\subsection{Initial Framework Design}

\begin{figure}[h]
\centering
\includegraphics[scale=0.5]{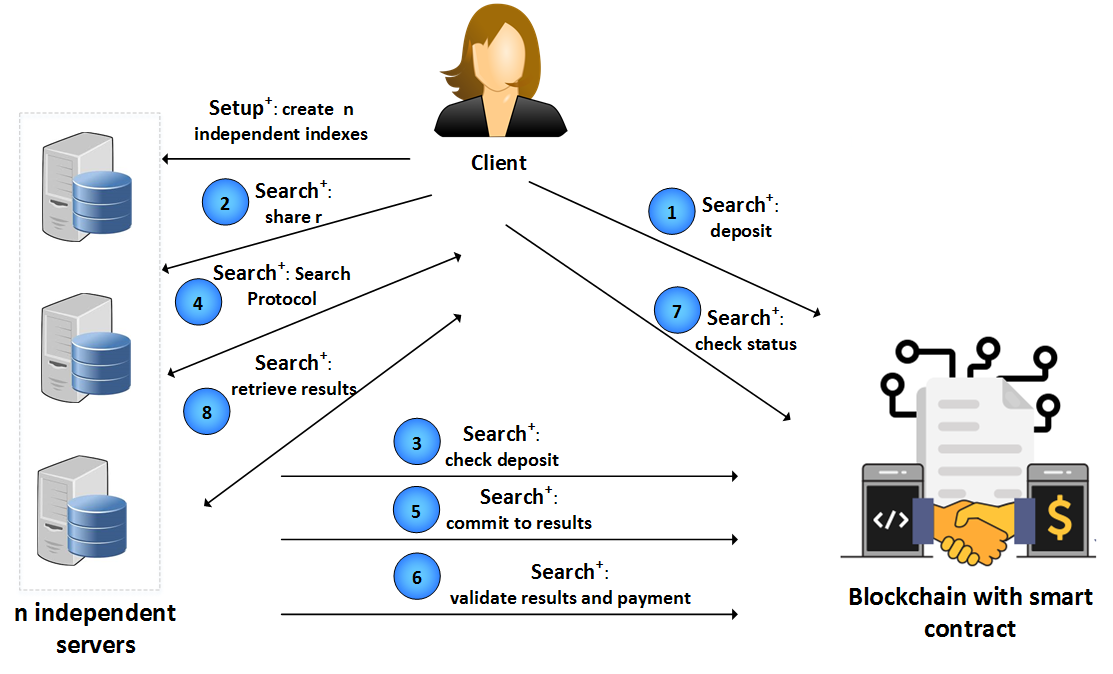}
\caption{Initial Design}
\label{fig:initial}
\end{figure}

In this design, the client needs to choose multiple servers for the sake of facilitating fairness while limiting information leakage to the Blockchain. Suppose there is a symmetric searchable encryption scheme $(\mathsf{Setup}, \mathsf{Search})$, which can be abstracted in the manner of Section \ref{pre}. Shown in Figure \ref{fig:initial}, leveraging on a Blockchain, the new construction consists of two stages $(\mathsf{Setup}^{\dag}, \mathsf{Search}^{\dag})$.

\begin{itemize}
\item $\mathsf{Setup}^{\dag}$ Stage: The client chooses $n$ servers which will not collude all together by assumption. The client then runs $\mathsf{Setup}$ $n$ times to generate $n$ independent encrypted indexes for its database. Finally, the client stores the indexes on the servers, where every server receives a unique index.

\vspace{0.1cm}

\item $\mathsf{Search}^{\dag}$ Stage: Given any keyword $w$, the search operation goes with the following phases.

    \begin{enumerate}

    \item \emph{Request} phase: The client deposits a certain amount of money on the Blockchain. The money should cover the cost of search operations of the servers and the operational cost of Blockchain for the whole workflow (i.e. this and next phases). Simultaneously, the client chooses a random number $r$ and sends it to all the servers to initiate a search operation.

    \vspace{0.1cm}

    \item \emph{Search} phase: Every server verifies that sufficient money has been deposited on the Blockchain. If the verification passes, it runs the $\mathsf{Search}$ protocol with the client. Regarding the abstraction of $\mathsf{Search}$ in Section \ref{pre}, the server skips Step 3, and instead it does the following.
        \begin{enumerate}
        \item Compute a hash value of the form: $\mathsf{H}(\mathcal{ID}_w||r)$, where $\mathcal{ID}_w$ contains the matched file identifiers and $\mathsf{H}$ is a cryptographic hash function.
        \item Run a commitment scheme (e.g. that from \cite{Pedersen}) to generate a commitment $commit$ for $\mathsf{H}(\mathcal{ID}_w||r)$.
        \item Store $commit$ on the Blockchain.
        \end{enumerate}

    \vspace{0.1cm}

    \item \emph{Validation} phase: Every server checks that all other servers have sent their commitments to the Blockchain. If so, it sends its key, which is related to the commitment scheme, to the Blockchain. The smart contract opens all the commitments with the corresponding keys and stores $\mathsf{H}(\mathcal{ID}_w||r)$ on the Blockchain. If all the opened results are the same, then the smart contract makes a payment using the deposited money to every server. Otherwise, the smart contract stops and leaves the client and servers to solve the dispute offline.

    \vspace{0.1cm}

    \item \emph{Retrieval} phase: If payments have been made, the client requests all the servers to send back the file identifiers $\mathcal{ID}_w$. It can validate the received $\mathcal{ID}_w$ based on the hash value $\mathsf{H}(\mathcal{ID}_w||r)$ and the random number $r$.

    \end{enumerate}

\end{itemize}

It is easy to check that if the client and servers are semi-honest, then the searchable encryption solution will work properly. Comparing with the solution from Section \ref{sec:huscheme}, it is clear that the Blockchain has very light involvement here: mainly storing deposit and validating the hashed search results, i.e. the hash values $\mathsf{H}(\mathcal{ID}_w||r)$. Next, we evaluate the overall security of this design by answering the following questions.

\emph{How has the privacy guarantee of the original searchable encryption scheme been affected?} From the perspective of an individual server, it is easy to see that adapting a searchable encryption scheme to the new framework does not affect the privacy properties of the original scheme. In another word, the information leakage to an individual server remains the same. When several servers collude, it becomes quite tricky, at least for those schemes which can only be proven secure in the indistinguishability-based security models. Nevertheless, if the underlying searchable encryption scheme adopts a simulation-based security definition, e.g. \cite{curtmola:1}, we conjecture that the collusion of all servers leaks the same amount of information as in the case of a single server. A rigorous proof of this conjecture will be an interesting future work.

\emph{How does the Blockchain affect the privacy and other security properties?}  As to data storage related to the index and search results, the Blockchain only stores $\mathsf{H}(ID||r)$ which is a random value if $\mathsf{H}$ is modelled as a random oracle. Therefore, the Blockchain does not affect the privacy guarantee regarding all potential attackers: one server, multiple servers, and even all servers. To this end, whether the Blockchain is permissioned or permissionless does not have any impact on privacy. 

\emph{How has the fairness property been achieved?} We analyse the fairness property from perspectives of the client and servers, respectively.
\begin{itemize}
\item It is easy to check that if at least one server is semi-honest then any misbehaved server will be detected and no payment will be made. Here, the misbehaviour mainly means that a server does not commit to the right $\mathcal{ID}_w$. It is left as an offline task to figure out who are the cheated servers and how to compensate for the semi-honest ones.
\item \emph{Let's assume the indexes have been generated faithfully by the client in the $\mathsf{Setup}^{\dag}$ Stage and the random number $r$ and trapdoors issued to the servers are properly generated in the $\mathsf{Search}^{\dag}$ Stage}. If all servers carry out the search honestly then they will be paid by the smart contract in the \emph{Validation} Phase, regardless of the client's decision at that point. However, if the client is malicious and deviates from the  protocol specification, e.g. issuing a wrong trapdoor to a specific server, then even if a server is honest it might not be paid.
\end{itemize}

Regarding the Blockchain platform, whether it is permissioned or permissionless is not a concern. Instead, one important criteria is the time duration that a new block gets added to the Blockchain, which determines the execution time of the $\mathsf{Search}^{\dag}$ Stage. Another criteria is the overall security of the platform, e.g. its capability to defend DDoS attacks and its tolerance to compromised miners. Comparing with the solutions from \cite{chen19,Hu}, even if the underlying Blockchain platform is broken down, the smart contract can be easily deployed on a new Blockchain platform to sustain the operations while the other simpler option is to continue without the operations depending on the Blockchain to achieve privacy protection only.

In summary, the above design does not provide a full-fledged solution for fairness because offline operations need to be carried out to solve the dispute and compensate for all honest behaviours. As a lesson, it tells us that employing multiple servers does not necessarily lead to a straightforward achievement of fairness. This is due to the fact that we do not want the client and servers to make their operations transparent and publicly verifiable for the sake of privacy protection, in contrast to the analysed Blockchain-based solutions.

\subsection{Improved Framework Design}

The initial design from the previous subsection has two main caveats. One is about the security guarantee when several servers collude. The other is that the fairness property relies on the assumption that at least one of the servers is semi-honest and will not collude with the rest. It is easy to check that if all the servers collude then they can fake the $\mathcal{ID}_w$ together so that the validation by the Blockchain will not detect the cheating. In practice, it might be difficult to decide how many servers should be chosen to fulfill the assumption, i.e. setting the $n$ parameter. If we simply assume all the servers are semi-honest, a common assumption in many papers, then both caveats will not be an issue. However, it is desirable to get rid of them technically. An additional minor caveat is that the client might misbehave in the protocol, e.g. sends a wrong index or trapdoor to a specific server, so that fairness will not be achieved even if all servers are semi-honest. Although this is very unlikely to occur for a rational client in practice, but it remains as a potential concern.

\begin{figure}[h]
\centering
\includegraphics[scale=0.5]{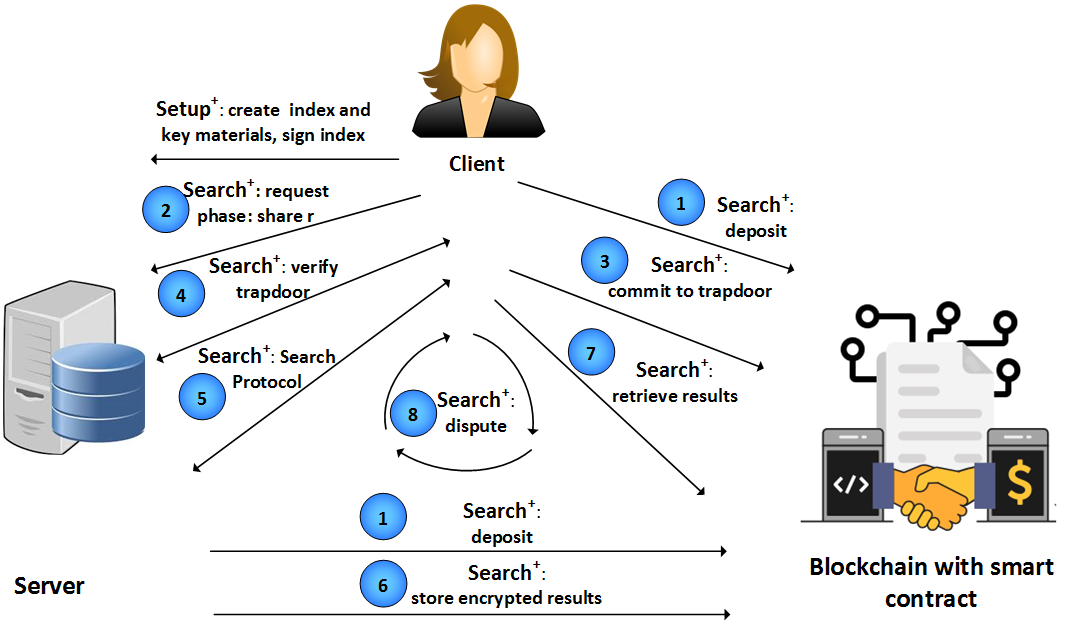}
\caption{Improved Design}
\label{fig:improved}
\end{figure}

To eliminate these caveats with technical countermeasures, we propose an improved design, shown in Figure \ref{fig:improved}. Since we do not intend to improve or affect the security guarantee of the original searchable encryption scheme, this improved design aims at achieving the fairness property while avoiding the caveats in the initial design. To this end, we let the client sign the encrypted index and the trapdoors so that no entity can misbehave and deny its misbehaviour. Public-key encryption and zero-knowledge proofs are employed to preserve the privacy guarantee of the original scheme. We require also the server to deposit money on the Blockchain to deter its cheating incentives.

Suppose there is a symmetric searchable encryption scheme $(\mathsf{Setup}, \mathsf{Search})$, which can be abstracted in the manner of Section \ref{pre}. The improved construction has two new stages $(\mathsf{Setup}^{\dag}, \mathsf{Search}^{\dag})$.

\begin{enumerate}
\item $\mathsf{Setup}^{\dag}$ Stage: The client runs $\mathsf{Setup}$ to generate a searchable index for its database $\mathcal{DB}$, with the following deviations.
        \begin{itemize}
            \item For each keyword $w \in \mathcal{W}$, besides the associated file identifier set $\mathcal{ID}_w$, the client generates a private key $K_w$ and adds an additional virtual identifier $id_w^{*}=\mathsf{H}(K_w||\mathcal{ID}_w)$ where $\mathsf{H}$ is a cryptographic hash function and $\mathcal{ID}_w$ denotes the concatenation of file identifiers in lex order. We further assume that $id_w^{*}$ can be easily distinguished from those identifiers in $\mathcal{ID}_w$.


            \item The client chooses an EU-CMA (Existential Unforgeability under a Chosen Message Attack) secure signature scheme $(\mathsf{KeyGen}_s, \mathsf{Sign}, \mathsf{Verify})$ and runs $\mathsf{KeyGen}_s$ to generate a sign/verification key pair $(sk_s, vk_s)$. It also chooses an IND-CPA (Indistinguishability under chosen-plaintext attack) secure public key encryption scheme $(\mathsf{KeyGen}_e, \mathsf{Enc}, \mathsf{Dec})$ and runs $\mathsf{KeyGen}_e$ to generate an encryption/decryption key pair $(pk_e, sk_e)$.
        \end{itemize}
    Besides the required activities in the original $\mathsf{Setup}$ procedure, the client stores $K_w$ $(w \in \mathcal{W})$, $sk_s$, and $sk_e$ locally, and stores the public keys on the Blockchain. The client also generates a signature $sig_I$ for the encrypted index and stores it on the Blockchain.

    We assume the smart contract on the Blockchain platform has been deployed with the following functions.
        \begin{itemize}
            \item $\mathsf{Deposit}$: the client or the server can call this function to deposit money which can be used to make payments.
            \item $\mathsf{Dispute}$: the client can call this function to resolve cheating activities.
            \item $\mathsf{SearchOK}$: the client explicitly validates the search result and makes a payment.
        \end{itemize}


\item $\mathsf{Search}^{\dag}$ Stage: Given any keyword $w$, the search operation goes with the following phases.

    \begin{enumerate}
        \item \emph{Request} phase: Both the client and the server deposit a certain amount of money on the Blockchain, by invoking the $\mathsf{Deposit}$ function of the smart contract. The client's money should cover the cost of search operation of the server, the operational cost of Blockchain for the whole workflow (i.e. this and next steps excluding dispute resolution), and the cost of dispute resolution function $\mathsf{Dispute}$. While the server's money should cover the operational cost of Blockchain for the whole workflow, the cost of dispute resolution function $\mathsf{Dispute}$, plus a pre-agreed amount for punishing its potential cheating behaviour.

            At the end of this phase, the smart contract verifies the deposits and indicates the client and the server to proceed or not.


        \item \emph{Search} Phase: The server runs the $\mathsf{Search}$ protocol with the client. Referring to the abstraction of $\mathsf{Search}$ in Section \ref{pre}, we add the following extra operations.

            \begin{itemize}
                \item At the end of Step 1, the client chooses a random number $r$ and generates a signature $sig_w = \mathsf{Sign}(\mathsf{H}(r||T_w), sk_s)$. It stores $sig_w$ on the Blockchain and shares $r$ with the server.


                \item At the beginning of Step 2, the server retrieves the $sig_w$ from the Blockchain and verifies it according to the received trapdoor and random number $r$ by the $\mathsf{Verify}$ algorithm. It also verifies the signature $sig_I$ for the encrypted index.  If the verifications pass, it continues the operations; otherwise it aborts.


                \item In Step 3, after recovering the file identifiers $id_w^*, \mathcal{ID}_w$, the server  encrypts them with $pk_e$ and stores the ciphertext $C_w$ on the Blockchain.
            \end{itemize}


        \item \emph{Retrieval \& Validation} Phase: Before going to the details, we define two actions first.
            \begin{itemize}
            \item \textbf{Action type-1}: the smart contract (1) makes a payment to the server; (2) return the server's deposit back.
            \item \textbf{Action type-2}: the smart contract (1) pays the server's deposit to the client by deducing the amount for smart contract execution and the server's deposit for dispute resolution function $\mathsf{Dispute}$.
            \end{itemize}

        In this phase, the client retrieves the ciphertext $C_w$ and obtains the plaintext: $id_w^*,\mathcal{ID}_w$. Then the client verifies the search results by checking $id_w^{*}=\mathsf{H}(K_w||\mathcal{ID}_w)$.
            \begin{itemize}
                \item If the verification passes, the client invokes the $\mathsf{SearchOK}$ function which will carry out \textbf{Action type-1} and return the client's deposit for dispute resolution function $\mathsf{Dispute}$.


                \item If the verification fails, the client invokes the $\mathsf{Dispute}$ function, by providing $id_w^*,\mathcal{ID}_w$ and a zero-knowledge proof $\mathcal{P}_1$ showing that $id_w^*,\mathcal{ID}_w$ are the plaintext of the ciphertext $C_w$; the server is required to provide $r$, $T_w$. The $\mathsf{Dispute}$ function does the following.
                    \begin{enumerate}
                    \item Verify the signature $sig_w$ based on the received $r$ and $T_w$. If the verification passes, it continues; otherwise it carries out \textbf{Action type-2} in favor of the client.


                    \item Request the server to upload a copy of encrypted index, denoted as $\mathcal{I}$, and verify its signature $sig_I$ which has been stored on the Blockchain. If the verification passes, it continues; otherwise it carries out \textbf{Action type-2} in favor of the client.


                    \item Verify the proof $\mathcal{P}_1$. If the verification passes, it continues; otherwise it carries out \textbf{Action type-1} in favor of the server.


                    \item Execute the $\mathsf{Search}$ procedure with $T_w$ to obtain ${id_w^*}',\mathcal{ID}_w'$. If these values are different from $id_w^*,\mathcal{ID}_w$, it carries out \textbf{Action type-2} in favor of the client. Otherwise, it carries out \textbf{Action type-1} in favor of the server.
                    \end{enumerate}

            \end{itemize}
        If the $\mathsf{Dispute}$ function is not invoked, the smart contract carries out \textbf{Action type-1} and returns the client's deposit for dispute resolution function $\mathsf{Dispute}$.
\end{enumerate}
\end{enumerate}

It is easy to check that if the client and the server are semi-honest, then the searchable encryption solution will work properly. Unless there is a dispute to be resolved, the Blockchain has very light involvement in the new solution: mainly storing deposits, public keys, the signatures, encrypted search results, and so on. As to setting up of the client's signature scheme and encryption scheme in the $\mathsf{Setup}^{\dag}$ Stage, EU-CMA security is necessary to prevent other entities (e.g. the server) from forging the client's signature, while IND-CPA is adequate to protect the confidentiality of the encrypted data due to the fact that no outsider attacker is allowed to commit ciphertext to the Blockchain and get access to decryption oracle (see our unilateral integrity assumption made in the beginning of this section).

Next, we evaluate the overall security of this design by answering the same questions as those for the initial design.

\emph{How has the privacy guarantee of the original searchable encryption scheme been affected?} The main change to the original searchable encryption scheme is adding the virtual identifier $id_w^{*}$ for every keyword, and this clearly does not change any privacy guarantee of the scheme. From the view of the server, the Blockchain does not provide any new information about the trapdoor and encrypted index. Therefore, the new design leaks exactly the same amount of information to the server as in the original scheme.

\emph{How does the Blockchain affect the privacy and other security properties?}  Due to the fact that the Blockchain only stores the signatures, encrypted search results and other public information, therefore it does not affect security guarantee of the original searchable encryption scheme, regardless the category of the underlying Blockchain. When a dispute occurs, it will expose more information such as encrypted index and trapdoor. However, this will not cause any privacy problem, because it does not amplify the privacy concerns as in the case of solutions analysed in Section \ref{sec:dltsol}, due to the fact that only the encrypted index and one trapdoor is made public on the Blockchain.

\emph{How has the fairness property been achieved?} Compared with the initial design,  fairness is guaranteed without relying out any assumption on the client.
\begin{itemize}
\item We first analyse the fairness property for the client. Note that the added virtual identifiers $id_w^{*}$ in the $\mathsf{Setup}^{\dag}$ Stage is a HMAC (hash-based message authentication code) for the file identifiers associated with the keyword $w$. Therefore, after recovering $id_w^*, \mathcal{ID}_w$ in the \emph{Retrieval \& Validation} Phase, the client can determine whether the result is correct or not by verifying this value. If it is not correct, then the server must have misbehaved because applying legitimate $T_w$ to legitimate $\mathcal{I}$ will result in the correct $id_w^*, \mathcal{ID}_w$. With respect to the dispute resolution procedure in the \emph{Retrieval \& Validation} Phase, the server will be punished in either step i, ii, or iv.

\item We then analyse the fairness property for the server. For any search query, if the server honestly carries out the operation, then the following will hold: (i) a random number $r$, a trapdoor $T_w$, and a valid signature $sig_w$ stored on the Blockchain; (ii) the searchable index $\mathcal{I}$ is the original one with a valid signature $sig_I$ stored on the Blockchain;  (iii) a ciphertext $C_w$ stored on the Blockchain where the plaintext $id_w^*, \mathcal{ID}_w$ are the results of applying $T_w$ to $\mathcal{I}$. With respect to the definition of the \emph{Retrieval \& Validation} Phase, it is straightforward to verify that the server will receive a payment for its work.
\end{itemize}

As a quick remark, this improved solution incurs less cost than the initial solution, which multiple servers need to perform search operations in parallel.

\section{Conclusion}
\label{con}

In this paper, we analysed two Blockchain-based searchable encryption solutions and identified a number of issues with respect to economical, security and privacy issues. We demonstrated that Blockchain is not a silver bullet that can be used straightforwardly to solve fairness issues in reality. Based on the analysis, we presented two Blockchain-enabled frameworks which can be applied to most existing symmetric searchable encryption schemes to achieve fairness while preserving the original privacy guarantees. We provided corresponding analysis to the new designs, and showed that the improved construction achieves the same level of fairness as the existing Blockchain-based solutions without suffering their privacy problems. It was also shown that, as long as there is no dispute, the overhead of involving the Blockchain is very low. As an immediate next step, we plan  to implement the improved framework and demonstrate its performances with respect to concrete searchable encryption schemes and Blockchain platforms.

\bibliography{dlt}
\bibliographystyle{plain}

\end{document}